\documentclass[10pt,german]{article}
\usepackage{a4}

\author{Boris K\" ors\footnotemark[1],\\ Michael G. Schmidt\footnotemark[2]} 

\date{\small{Institut f\" ur Theoretische Physik, Philosophenweg 16, D-69120 Heidelberg, Germany}}

\title{The effective two-loop Euler-Heisenberg action for scalar and spinor QED in a general constant background field}

\begin{document}

\newcommand{\beq}{\begin{equation}}
\newcommand{\eeq}{\end{equation}}
\newcommand{\beqn}{\begin{eqnarray}}
\newcommand{\eeqn}{\end{eqnarray}}
\newcommand{\wf}{Worldline formalism}
\newcommand{\G}{G_{Bab}}
\newcommand{\g}{\dot{G}_{Bab}}
\newcommand{\Gv}{G^v_{Bab}}
\newcommand{\Gw}{G^w_{Bab}}
\newcommand{\gv}{\dot{G}_{Bab}^v}
\newcommand{\gw}{\dot{G}_{Bab}^w}
\newcommand{\tr}{{\mbox{tr}}}
\newcommand{\sign}{{\mbox{sign}}}
\newcommand{\dt}{{\mbox{det}}}

\maketitle
\thispagestyle{empty}

\begin{abstract}
Using the \wf\ of QED we compute the two-loop effective action induced by a charged scalar, respectively  spinor particle in a general constant electromagnetic field.  
\end{abstract}

\renewcommand{\thefootnote}{\fnsymbol{footnote}}
\footnotetext[1]{E-mail: B.Koers@Thphys.Uni-Heidelberg.de.}
\footnotetext[2]{E-mail: M.G.Schmidt@Thphys.Uni-Heidelberg.de.}

\vspace{1cm}

\section{Introduction}

The Euler-Heisenberg one-loop effective action, i.e. the effective action induced by an electron loop in a constant electromagnetic field, was one of the first problems to be adressed in the framework of Quantum Field Theory\cite{EH,SCH}. In the language of conventional Quantum Electrodynamics (QED) all corrections caused by a single electron loop in a constant electromagnetic field were summarized by regarding this field as a classical background field and introducing an extra term into the electron propagator. Later on these calculations were generalized to two-loop level\cite{RIT,RIT2,DR}. \\

The \wf\ was discussed recently\cite{POL,STR,SS3} as a method inspired by analogies to string theory in the limit of infinite string tension\cite{BK}. It allows the summation of whole classes of Feynman graphs and therefore highly reduces the effort of computing loop-corrections.  Its power has been demonstrated reproducing field theoretical results particularly in one- and two-loop QED\cite{SS1,SS2}. An application of the \wf\ to the Euler-Heisenberg problem turns out to be natural, as the approach chosen in standard field theory in this case is already somewhat similar to the Worldline philosophy. Indeed it reduces to performing Gaussian path integrals in the space of periodic and antiperiodic functions on the circle. In this publication we give a generalization and completion of earlier work by D. Fliegner, M. Reuter, C. Schubert and one of the authors\cite{CEF,DIMR}. While they had only treated purely magnetic fields before, we now cover general constant electromagnetic field configurations. We again employ the dimensional regularization scheme, that had been used to overcome technical problems of proper time regularization. After a short introduction which will provide the master-formulas for the two-loop corrections we shall separately discuss scalar and spinor QED.\\

 Such results are of immediate use e.g. for testing quantum corrections to the polarization tensor of the photon by means of a laser beam in an intense electromagnetic field \cite{EXP} or for computing corrections to the famous Casimir energy density between two parallel plates\cite{FR}. Using the \wf\ it has also recently been possible to clarify discrepancies between different results for amplitudes of photon splitting in a background field\cite{AS}.

\section{The \wf\ of QED in a general constant electromagnetic background field}

Exploring the Dirac-action of a massive spinor-particle which couples through the covariant derivative $D_\mu \equiv \partial_\mu -ieA_\mu$ to an abelian gauge field $A_\mu$ in the second order formalism, one can express the effective action induced by a Dirac-particle loop in an electromagnetic background field with potential $A_\mu$ and field strength tensor $F_{\mu \nu}$ by: 
\beqn \label{einlo}
\Gamma[A]= -2\int_{0}^{\infty}{\frac{dT}{T} e^{-m^2T} {\cal D}x{\cal D}\psi}\ \exp \left( -\int_{0}^{T}{d\tau \left( \frac{\dot{x}^2}{4} +\frac{1}{2} \psi \dot{\psi} +ieA_\mu \dot{x}^\mu -ie \psi^\mu F_{\mu \nu} \psi^\nu\right)} \right).
\eeqn
The path integrals are performed over scalar fields $x(\tau)$ using periodic boundary conditions $x(0)=x(T)$ and their Grassmannian superpartners $\psi(\tau)$ with antiperiodic boundary conditions $\psi(0)=-\psi(T)$. The ``centre of mass'' coordinate 
\beq
x_0 \equiv \int_0^T{x(\tau) d\tau}
\eeq
of the scalar fields will then be subtracted, which results in a delta function for the overall momentum conservation. The Worldline Lagrangian is invariant under the world-sheet supersymmetry \cite{BVH}
\beqn
\delta_\epsilon x^\mu = -\frac{1}{2} \epsilon \psi^\mu, \quad \delta_\epsilon \psi^\mu = \epsilon \dot{x}^\mu
\eeqn
and the interaction part of $\Gamma[A]$ can therefore be written as a supersymmetric Wilson-loop. For this the two-loop correction again is known \cite{SS1} and given by the double contour integral insertion of
\beqn \label{zweiloopeinf}
& & \frac{e^2}{2} \frac{\Gamma \left( d/2-1 \right)}{4\pi^{d/2}} \int_{0}^{T}{d\tau_a d\theta_a \int_{0}^{T}{d\tau_b d\theta_b \frac{D X^\mu_a D X_{\mu b}}{(X_a-X_b)^{d-2}}}} \\
&=&  \frac{e^2}{2} \int_{0}^{T}{d\tau_a d\theta_a \int_{0}^{T}{d\tau_b\ d\theta_b \ D X^\mu_a D X_{\mu b} \int_{0}^{\infty}{d\bar{T}\ (4\pi T)^{-d/2} \exp \left( -\frac{(X^\mu_a-X^\mu_b)^2}{4\bar{T}} \right) }}} \nonumber 
\eeqn
into the one-loop path integral (\ref{einlo}). For brevity we introduced the superfield notation
\beqn
X^\mu_a(\hat{\tau}) &\equiv & X^\mu_a(\tau,\theta) \equiv x^\mu(\tau_a) + \sqrt{2} \ \theta \psi^\mu (\tau_a), \\
D &\equiv & \frac{\partial}{\partial \theta} -\theta \frac{\partial}{\partial \tau} \nonumber
\eeqn
with Grassmannian variables $\theta_{a,b}$, but we shall return to component field representation later. For simplification of further notation we also define the operator $B_{ab}$ by
\beq
(x(\tau_a)-x(\tau_b))^2 \equiv \int_{0}^{T}{d\tau_1 \int_{0}^{T}{d\tau_2 \left( x(\tau_1) B_{ab} x(\tau_2) \right) }}.
\eeq
Assuming now the gauge field to be a constant, classical background field, we can employ the Fock-Schwinger gauge $A_\mu= \frac{1}{2} F_{\mu \nu} x^\nu$. Thereby all terms in the exponential become bilinear and the evaluation of the two-loop effective action reduces to the computation of Gaussian integrals, i.e. the computation of the functional determinants of the the two bilinear operators considering their respective boundary conditions. To obtain explicit results one can for instance use the spectral representation in the Fourier basis \cite{CEF}
\beqn \label{determ}
{\mbox{Det}}_P \left( -\frac{d^2}{d\tau^2} +2ieF\frac{d}{d\tau} +\frac{B_{ab}}{\bar{T}} \right) &=& (4\pi T)^d \det \left( \frac{\sin(eFT)}{eFT} \right) \det \left( {\bf I}-\frac{C_{ab}}{\bar{T}} \right) \nonumber \\
{\mbox{Det}}_A \left( {\bf I}-2ieF \left( \frac{d}{d\tau} \right)^{-1} \right) &=& \det \left( \cos(eFT) \right). 
\eeqn
The operator $C_{ab}$ will be defined in (\ref{cab}). We now only remain with performing the ordinary determinants over Lorentz-indices. For this we shall have to specify the Lorentz-frame we shall be working in in the following.\\

For scalar QED, the field theory of spinless, massive point particles, coupling to an abelian gauge field, as expressed by the euclidean action
\beq 
S_{{\mbox{\scriptsize scal}}}= \phi^\dag (-D^2+m^2) \phi, 
\eeq
one finds in the \wf\ completely analogously effective actions for the one- and two-loop level. The corresponding formulas are obtained from (\ref{einlo}) and (\ref{zweiloopeinf}), rewritten in the component field representation, by erasing all terms containing $\psi$, as well as the prefactor $-2$. Alternatively stated in reversed order this means, that spinor QED is obtained from scalar QED by substituting $x(\tau)$ by a supervariable coordinate $X(\hat{\tau})$. In the following two chapters we shall now explicitly compute the expressions we have got and find their functional and asymptotic dependence on the field strengths of the electric and magnetic background fields, always using dimensional regularization. This enables a direct comparison of these results of the \wf\ to similar calculations of ordinary QED.

\section{Scalar Quantum Electrodynamics}
\label{skalar}   

We shall first treat the simpler case of scalar theory and afterwards find its generalization to spinning particles. The starting point is the two-loop correction to the Euler-Heisenberg Lagrangian in the \wf\ of scalar QED in a constant electromagnetic background field. Using (\ref{determ}) in (\ref{einlo}) and (\ref{zweiloopeinf}),we find \cite{CEF}:
\beqn \label{master}
{\cal L}^{(2)}_{{\mbox{\scriptsize scal}}}[F] &=& -\frac{e^2}{2(4\pi)^{d}} \int_{0}^{\infty}{\frac{dT}{T^{d/2+1}}e^{-m^2T}} \int_{0}^{\infty}{d\bar{T}} \int_{0}^{T}{d\tau_a} \int_{0}^{T}{d\tau_b} \nonumber \\
& & \times \dt^{-1/2} \left( \frac{\sin(eFT)}{eFT} \right) \dt^{-1/2} \left( \bar{T}-\frac{1}{2} {\cal C}_{ab} \right) \langle \dot{x}_a \dot{x}_b \rangle.
\eeqn
The remaining contraction of the bosonic variables $\dot{x}_{a,b} \equiv \dot{x}(\tau_{a,b})$ can be rewritten in terms of Green's functions:
\beq
\langle \dot{x_a} \dot{x_b} \rangle = \tr\left( \ddot{{\cal G}}_{Bab}+\frac{1}{2} \frac{(\dot{\cal G}_{Baa}-\dot{\cal G}_{Bab})(\dot{\cal G}_{Bab}-\dot{\cal G}_{Bbb})}{\bar{T}-\frac{1}{2} {\cal C}_{ab}} \right).
\eeq
We used the modified Worldline Green's function
\beqn
{\cal G}_{Bab} \equiv {\cal G}_{B}(\tau_a,\tau_b) &\equiv& \langle \tau_a \mid \left( \frac{1}{2} \left( \frac{d^2}{d\tau^2}-2ieF\frac{d}{d\tau} \right) \right)^{-1} \mid \tau_b \rangle \\
&=&\frac{1}{2(eF)^2} \left( \frac{eF}{\sin(eFT)} e^{-ieFT\g}+
ieF\g -\frac{1}{T} \right) \nonumber
\eeqn
and
\beqn \label{cab}
{\cal C}_{ab} &\equiv & {\cal G}_{Baa} -{\cal G}_{Bab} -{\cal G}_{Bba} +{\cal G}_{Bbb} \\
&=& \frac{ \cos(eFT) -\cos(eFT\g)}{eFT\ \sin(eFT)}. \nonumber
\eeqn
The ordinary Worldline Green's function inverts the operator $\frac{1}{2} \partial_\tau^2$ on a circle of radius $T$. Up to an irrelevant constant it is given by
\beq
G_{Bab} \equiv G_{B}(\tau_a,\tau_b) \equiv \mid \tau_a-\tau_b \mid - \frac{(\tau_a-\tau_b)^2}{T},
\eeq
and its derivative with respect to the first variable is:
\beq
\dot{G}_{Bab}=\sign(\tau_a -\tau_b)-2\frac{(\tau_a-\tau_b)}{T}.
\eeq
After a partial integration in $\tau_a$ (\ref{master}) can be rearranged in a form, in which the divergencies appearing in the integrations over $T$ and $\tau_{a,b}$ are combined differently and in a very suitable manner. This will be of particular use when we regularize the integrations by introducing a new linear combination of the two expressions:
\beqn \label{maspar}
\hat{\cal L}^{(2)}_{{\mbox{\scriptsize scal}}}[F] &=& -\frac{e^2}{2(4\pi)^{d}} \int_{0}^{\infty}{\frac{dT}{T^{d/2+1}}e^{-m^2T}} \int_{0}^{\infty}{d\bar{T}} \int_{0}^{T}{d\tau_a} \int_{0}^{T}{d\tau_b} \\
& & \times \dt^{-1/2} \left( \frac{\sin(eFT)}{eFT} \right) \dt^{-1/2} \left( \bar{T}-\frac{1}{2} {\cal C}_{ab} \right) \nonumber \\
& & \times \frac{1}{2} \left( \tr(\dot{\cal G}_{Bab}) \tr\left( \frac{\dot{\cal G}_{Bab}}{\bar{T}-\frac{1}{2} {\cal C}_{ab}} \right) +\tr\left( \frac{(\dot{\cal G}_{Baa}-\dot{\cal G}_{Bab})(\dot{\cal G}_{Bab}-\dot{\cal G}_{Bbb})}{\bar{T}-\frac{1}{2} {\cal C}_{ab}} \right) \right). \nonumber
\eeqn
Now all quantities in the Lagrangian are written as functions only depending on the Schwinger proper time (SPT) variables $T$, $\bar{T}$ and $\tau_{a,b}$, the Worldline Green's function $G$ as well as the field strength tensor $F$ of the electromagnetic field.\\

It is well known that for any such field strength tensor there either exists a Lorentz-frame, in which the electric and magnetic fields are parallel and their magnitudes $\epsilon$ and $\eta$ in this frame are relativistic invariants of the field, or they are perpendicular in any frame\cite{LAN}. In the latter case a Lorentz transformation can be used to eliminate one of the fields, so we only have to deal with the first case. Therefore $F$ takes on a very simple form where only two symplectic block elements are non zero, so that the determinants just factorize and the traces split into sums of the different block-traces. One can further diagonalize $F$ by the unitary transformation 
\beqn
U \equiv \frac{1}{\sqrt{2}} \left( 
\begin{array}{cccc}
1 & i & 0 & 0 \\ i & 1 & 0 & 0 \\ 0 & 0 & 1 & i \\ 0 & 0 & i & 1 
\end{array} \right)
\eeqn
and read off the field strengths from the complex eigenvalues $a$ and $b$ of $U^\dagger F U$. To simplify the notation by symmetrizing all expressions we introduce the eigenvalues themselves as new variables:
\beq \label{epseta}
a\equiv \epsilon \qquad \mbox{and} \qquad b\equiv -i\eta.
\eeq
In the conventional scalar QED parameter integral expressions for the one- \cite{EH,SCH} and two-loop \cite{RIT,RIT2,DR} contributions to the Euler-Heisenberg Lagrangian are known for a general, constant, respectively a purely magnetic, constant background field. Using the above notations the unregularized  effective one-loop Euler-Heisenberg Lagrangian from standard QED is given by:
\beqn
{\cal L}_{{\mbox{\scriptsize scal}}}^{\mbox{\scriptsize QED}}[F] &=& {\cal L}^{(0)}_{{\mbox{\scriptsize scal}}}[F]+{\cal L}^{(1)}_{{\mbox{\scriptsize scal}}}[F] \\
&\equiv & \frac{a^2+b^2}{2}+\frac{1}{16\pi^2} \int_{0}^{\infty}{\frac{dT}{T^3} e^{-m^2T} \left( \frac{e^2abT^2}{\sinh(eaT)\sinh(ebT)} \right)}. \nonumber
\eeqn
Later on we shall need the subtracted version of the one-loop correction in dimensional regularization:
\beq
\bar{{\cal L}}^{(1)}_{{\mbox{\scriptsize scal}}}[F]=\frac{1}{(4\pi)^{d/2}} \int_{0}^{\infty}{\frac{dT}{T^{d/2+1}} e^{-m_0^2T} \left( \frac{e^2abT^2}{\sinh(eaT)\sinh(ebT)} +\frac{e^2(a^2+b^2)T^2}{6} -1 \right)}.
\eeq
Compared to the references given above, a Wick rotation of the integration variable was performed in both formulas. Regarding the antisymmetry and blockform of the field strength tensor, it is only necessary to know the trigonometric functions of some matrix $\sigma$ for the explicit computation of all functional expressions appearing in the Worldline formula (\ref{master}):
\beq
\sigma \equiv \left( \begin{array}{cc} 0 & 1 \\ -1 & 0 \end{array} \right).
\eeq
These are easily obtained by their power series expansions:
\beqn \label{sincos}
\sin(\sigma f) &=& \sigma \ \sinh(f), \nonumber \\
\cos(\sigma f) &=& I \ \cosh(f). 
\eeqn
Using the definitions
\beqn
I & \equiv & \left( \begin{array}{cc} 1 & 0  \\ 0 & 1 \end{array} \right), \nonumber \\
I_1 & \equiv & \left( \begin{array}{cc} I & 0 \\ 0 & 0 \end{array} \right), \nonumber \\
\sigma_1 & \equiv & \left( \begin{array}{cc} \sigma & 0 \\ 0 & 0 \end{array} \right),
\eeqn
and analogous expressions for $I_2$ and $\sigma_2$, $F$ reads
\beq
F=a\sigma_1 +b\sigma_2.
\eeq
It is to be heeded that powers of $F$ are evaluated in the Euclidean metric, otherwise $\sigma_2$ would always come with factors of $i$. Now we can freely employ (\ref{sincos}) to calculate the two-loop correction (\ref{master}). Futher introducing $v\equiv eaT$ and $w\equiv ebT$ we get:
\beqn \label{zwres}
{\cal G}_{Bab} & = & \frac{T}{2} \left( I_1 \left( \frac{\cosh(v\dot{G}_{Bab})}{v \sinh(v)} +\frac{1}{v^2} \right) -i\sigma_1 \left( \frac{\sinh(v\dot{G}_{Bab})}{v \sinh(v)}+\frac{\dot{G}_{Bab}}{v} \right) \right. \\
& & +\left. I_2 \left( \frac{\cosh(w\dot{G}_{Bab})}{w \sinh(w)} +\frac{1}{w^2} \right) -i\sigma_2 \left( \frac{\sinh(w\dot{G}_{Bab})}{w \sinh(w)}+\frac{\dot{G}_{Bab}}{w} \right) \right), \nonumber \\
\dot{\cal G}_{Bab} &=& I_1 \frac{\sinh(v \dot{G}_{Bab})}{\sinh(v)}
-i\sigma_1 \left( \frac{\cosh(v\dot{G}_{Bab})}{\sinh(v)} -\frac{1}{v} \right) + ( 1 \leftrightarrow 2, \quad v \leftrightarrow w ), \nonumber \\
\ddot{{\cal G}}_{Bab} &=& -I_1 \frac{2v}{T} \frac{\cosh(v\dot{G}_{Bab})}{\sinh(v)}+i\sigma_1 \frac{2v}{T} \frac{\sinh(v\dot{G}_{Bab})}{\sinh(v)}+(1 \leftrightarrow 2, \quad v \leftrightarrow w), \nonumber \\
\dot{\cal G}_{Baa} &=& i \cot(eFT)-\frac{i}{eFT} = i\sigma_1 \left( \frac{1}{v} -\coth(v) \right) +i\sigma_2 \left( \frac{1}{w} -\coth(w) \right), \nonumber \\
{\cal C}_{ab} & = & -I_1 \left( \frac{\cosh(v)-\cosh(v\dot{G}_{Bab})}{v \sinh(v)} \right) -I_2 \left( \frac{\cosh(w)-\cosh(w\dot{G}_{Bab})}{w \sinh(w)} \right).\nonumber
\eeqn
With some additional simplification of notation:
\beqn
G^v_{Bab} & \equiv & \frac{T}{2} \frac{\cosh(v)-\cosh(v\dot{G}_{Bab})}{v \sinh(v)}, \nonumber \\
\dot{G}_{Bab}^v & \equiv & \frac{\sinh(v\dot{G}_{Bab})}{\sinh(v)}, \nonumber \\
\gamma^v & \equiv & \frac{1}{\bar{T} +\Gv}, \nonumber \\
\gamma & \equiv & \frac{1}{\bar{T}+\G},
\eeqn
and similar definitions for $\Gw$, $\gw$ and $\gamma^w$, one can insert (\ref{zwres}) into (\ref{master}). We prefer to state the result immediately in dimensional regularization. Therefore we follow the procedures in \cite{DIMR} and continue the field strength tensor to $d=4+\epsilon$ dimensions by zero columns and rows, rescale to the unit circle according to $\tau_{a,b} \equiv u_{a,b}T$ and finally use the translation invariance on the circle to eliminate the $u_b$-integration. This leads to:
\beqn \label{greenfu}
\G & = & u_a(1-u_a), \nonumber \\
\g & = & 1-2u_a =\sqrt{1-4\G}.
\eeqn
The contributions of the $\epsilon$ additional dimensions are obtained exactly like the terms from vacuum dimensions in \cite{CEF} and \cite{DIMR} and before integrating they are of course of the order $\epsilon$ in the integrand. We now get:
\beqn \label{dety}
& & \dt^{-1/2} \left( \frac{\sin(eFT)}{eFT} (\bar{T}-\frac{1}{2} {\cal C}_{ab}) \right) = \frac{v}{\sinh(v)} \frac{w}{\sinh(w)} \gamma^v \gamma^w \gamma^{d/2-2}, \nonumber \\
& & \langle \dot{y}_{a} \dot{y}_{b} \rangle = (4-d)-4 \left( v \frac{\cosh(v \g)}{\sinh(v)} + w \frac{\cosh(w \g)}{\sinh(w)} \right) \nonumber\\ & & - \gamma^v (\g^{v2} +4v^2 \G^{v2})- \gamma^w (\g^{w2}+4w^2 \G^{w2})- \gamma \frac{d-4}{2} \g^2,
\eeqn
and for (\ref{maspar}) we also need:
\beqn \label{try}
& & \frac{1}{2} \left( \tr(\dot{\cal G}_{Bab}) \tr\left( \frac{\dot{\cal G}_{Bab}}{\bar{T}-\frac{1}{2} {\cal C}_{ab}} \right) + \tr\left( \frac{(\dot{\cal G}_{Baa}-\dot{\cal G}_{Bab})(\dot{\cal G}_{Bab}-\dot{\cal G}_{Bbb})}{\bar{T} - \frac{1}{2} {\cal C}_{ab}} \right) \right) = \nonumber \\
& & 2(\gv +\gw +\frac{d-4}{2} \g)(\gamma^v \gv +\gamma^w \gw +\gamma \frac{d-4}{2} \g) \nonumber \\ & & -\gamma^v (\g^{v2} +4v^2 \G^{v2}) -\gamma^w (\g^{w2} +4w^2 \G^{w2})- \gamma \frac{d-4}{2} \g^2.
\eeqn
The tadpole contribution proportional to $\delta(\tau_a -\tau_b)$ from $\ddot{\cal G}_{Bab}$ is vanishing in dimensional regularization and was already eliminated in (\ref{dety}). By inserting into (\ref{master}) and (\ref{maspar}) with (\ref{greenfu}), we obtain expressions for the integrands that exclusively depend on $T$, $\bar{T}$ and $u_a$. Now a discussion of the divergencies in the various integrations is possible and we do this by following carefully the steps of \cite{DIMR} again. Although the integration over $\bar{T}$ can also be performed in closed form at fixed and finite $T$ and $u_a$, we will prefer to expand the integrand in $T$ and $u_a$, the critical variables, and only then integrate the coefficients of the expansion term by term over $\bar{T}$. The necessary interchange of the integration over $\bar{T}$ with the differentiation with respect to $T$ is allowed by a rule of Leibniz. Writing the Worldline two-loop correction
\beq 
{\cal L}^{(2)}_{{\mbox{\scriptsize scal}}}[F]=-\frac{e^2}{2(4\pi)^d} \int_{0}^{\infty}{\frac{dT}{T^{d-1}} e^{-m^2T} \int_{0}^{\infty}{d\bar{T} \int_{0}^{1}{du_a\ I(T,\bar{T},u_a,d)}}},
\eeq
where we have called the rescaled variable again $\bar{T}$, we find that the integration over $T$ causes divergencies by those terms of the expansion of $I(T,\bar{T},u_a,d)$, that are constant or quadratic in $T$, as no odd powers are occurring at all. The analysis of the $u_a$-integration is more complicated. Here all powers of $\G$ that are negative in $d=4$ lead to divergencies. Particularly the power $\G^{1-d/2}$ contributes a $1/\epsilon$ pole:
\beq \label{geps}
\int_{0}^{1}{du_a \G^{1-d/2}}= \int_{0}^{1}{\frac{du_a}{(u_a(1-u_a))^{1+\epsilon/2}}}= B\left( -\frac{\epsilon}{2},-\frac{\epsilon}{2}\right) = -\frac{4}{\epsilon}+o(\epsilon),
\eeq
with the Euler-Betafunction $B(\alpha,\beta)\equiv \Gamma(\alpha) \Gamma(\beta)/ \Gamma(\alpha +\beta)$. Correspondingly the power $\G^{-d/2}$ implies another $1/\epsilon$ divergency. In fact, such terms are existing in the integrated coefficients of the expansion of $I(T,\bar{T},u_a,d)$ in powers of $T$:
\beqn
\int_{0}^{\infty}{d\bar{T}\ I(T,\bar{T},u_a,d)} &=& -\frac{d-2+8\G}{(d-2)\G^{d/2}} -\frac{v^2+w^2}{6d(d-2) \G^{d/2}} \\
& & \times \left( d(2-d)+2(d^2-6d+16)\G+16(d-14)\G^2 \right) +o(T^4). \nonumber
\eeqn
If we introduce a similar integrand $\hat{I}(T,\bar{T},u_a,d)$ for $\hat{\cal L}^{(2)}_{{\mbox{\scriptsize scal}}}[F]$ from (\ref{maspar}) and expand following the same procedures, we find
\beqn
\int_{0}^{\infty}{d\bar{T}\ \hat{I}(T,\bar{T},u_a,d)} &=& \frac{(d-1)(1-4\G)}{\G^{d/2}}-\frac{v^2+w^2}{6d \G^{d/2}} \\
& & \times \left( d(d-1)-(6d^2-22d+16)\G+8(d-2)(d-7)\G^2 \right)+o(T^4). \nonumber
\eeqn
We now recognize the possibility to arrive at an expression for the integrand, in which no orders $\G^{-d/2}$ are present in the expansion in powers of $T$, by using the linear combination:
\beq 
\frac{d-1}{d} {\cal L}_{{\mbox{\scriptsize scal}}}^{(2)}[F]+\frac{1}{d} \hat{\cal L}_{{\mbox{\scriptsize scal}}}^{(2)}[F].
\eeq
Terms of order $\G^{-d/2}$ cancel precisely, those in the constant contribution as well as those in the quadratic. Furthermore we got a cancellation of subdivergencies in the quadratic term, the one proportional to the Maxwell energy density $F^{\mu \nu}F_{\mu \nu}$, which is now of the order $\epsilon^0$. This results in a serious simplification of the regularization procedure, as we shall experience in the following. We again obtain an expansion:
\beqn \label{linkomb}
K(T,u_a,d) &\equiv & \int_{0}^{\infty}{d\bar{T} \left( \frac{d-1}{d} I(T,\bar{T},u_a,d) +\frac{1}{d} \hat{I}(T,\bar{T},u_a,d) \right) } \\ &=& K_{02}(T,u_a,d)+f(T,d)\G^{1-d/2}+o\left( T^4,\G^{2-d/2} \right), \nonumber
\eeqn
where now the divergencies are completely separated into
\pagebreak
\beqn \label{minsub}
K_{02}(T,u_a,d) &=& -4\frac{d-1}{d-2} \G^{1-d/2} -\frac{2(v^2+w^2)}{3d(d-2)}\\
& & \times \G^{1-d/2} (\G (2d^2-18d+4)-(d-4)(d-1)), \nonumber \\
f(T,d) &=& 4\frac{d-1}{d-2} -\frac{2(d-4)(d-1)}{3d(d-2)} (v^2+w^2) -\frac{8(d-1)}{d(d-2)} \frac{vw}{\sinh(v)\sinh(w)} \nonumber \\ 
& & \times \left( v \coth(v) + w \coth(w) + \frac{d-4}{2} \right). \nonumber
\eeqn
$K_{02}(T,u_a,d)$ carries all divergent contributions of the $T$-integration, while $f(T,d)$ is of fourth order in $T$ and therefore only divergent in the $u_a$-integration. After subtracting these two terms, the integral
\beqn
K(T,u_a,4)-K_{02}(T,u_a,4)-f(T,4)\G^{-1}
\eeqn
can easily be computed elementary and the remaining integrations over $T$ und $u_a$ stay finite. Therefore a set of subtraction terms (\ref{minsub}) is found and the Worldline two-loop correction to the Euler-Heisenberg Lagrangian from (\ref{master}) is regularized. If one further wants to obey the correct electron mass renormalization in the subtraction prescription, i.e. perform on shell subtraction, one has to adjust the finite part of the subtraction terms, which is determined by $f(T,d)$, in a way that the relation
\beq \label{onshbed}
\delta {\cal L}_{{\mbox{\scriptsize scal}}}^{(2)}[F]= \delta m_0^2 \frac{\partial}{\partial m_0^2} \bar{\cal L}^{(1)}_{{\mbox{\scriptsize scal}}}[F]
\eeq
with
\beq
\delta m_0^2 =\frac{\alpha m_0^2}{4\pi} \left( -\frac{6}{\epsilon}+7-3(\gamma-\ln(4\pi)) -\ln(m_0^2) \right)+o(\epsilon)
\eeq
holds. These conditions are derived in standard QFT, $\gamma$ is the usual Euler constant. Observing now that $f(T,d)$ satisfies
\beq
f(T,d)=\frac{8(d-1)}{d(d-2)} T^{1+d/2} \frac{d}{dT} \left( T^{-d/2} \left( \frac{vw}{\sinh(v)\sinh(w)} +\frac{v^2+w^2}{6} -1 \right) \right)
\eeq
one verifies by partial integration of the derivative term in $f(T,d)$ and the use of (\ref{geps}):
\beqn
\delta m_0^2 \frac{\partial \bar{\cal L}_{{\mbox{\scriptsize scal}}}^{(1)}[F]}{\partial m_0^2}&=&
-\frac{\alpha}{2(4\pi)^3} \int_{0}^{\infty}{\frac{dT}{T^{d-1}} e^{-m_0^2T}\int_{0}^{1}{du_a\ f(T,d)\ \G^{1-d/2}}} \\
& & +\frac{\alpha m_0^2}{(4\pi)^3} \int_0^{\infty}{\frac{dT}{T^2} e^{-m_0^2T} \left( -3(\gamma +\ln(m_0^2T))+\frac{3}{m_0^2T}+\frac{9}{2} \right) } \nonumber \\
& & \times \left( \frac{vw}{\sinh(v)\sinh(w)}+\frac{v^2+w^2}{6}-1 \right) +o(\epsilon). \nonumber
\eeqn
We now have the final result for the on shell renormalized two-loop Worldline correction to the Euler-Heisenberg Lagrangian for scalar QED:
\beqn \label{skalres}
\bar{\cal L}_{{\mbox{\scriptsize scal}}}^{(2)}[F] &=& -\frac{\alpha}{2(4\pi)^3} \int_{0}^{\infty}{\frac{dT}{T^3} e^{-m_0^2T}} \int_{0}^{1}{du_a \left( K(T,u_a,4)-K_{02}(T,u_a,4) \right)} -\delta m_0^2 \frac{\partial \bar{\cal L}_{{\mbox{\scriptsize scal}}}^{(1)}[F]}{\partial m_0^2} \nonumber \\
&=& -\frac{\alpha}{2(4\pi)^3} \int_{0}^{\infty}{\frac{dT}{T^3} e^{-m^2T}} \int_{0}^{1}{du_a \left( K(T,u_a,4)-K_{02}(T,u_a,4)-f(T,4)\G^{-1} \right)} \nonumber \\ 
& & +\frac{\alpha m^2}{(4\pi)^3}\int_{0}^{\infty}{\frac{dT}{T^2} e^{-m^2T} \left( \frac{e^2abT^2}{\sinh(eaT)\sinh(ebT)} +\frac{e^2(a^2+b^2)T^2}{6} -1 \right)} \nonumber \\ 
& & \times \left( -3(\gamma +\ln(m^2T))+\frac{3}{m^2T}+\frac{9}{2} \right).
\eeqn
Comparing to the result for a purely magnetic field in \cite{DIMR} a rather similar structure of terms is apparent. One only has to substitute in the sense that terms for the magnetic field are accompanied by electric ones. Still these substitutions are not as easy to be guessed in detail as the fairly simple structure of our result might suggest. Particularly the divergent subtraction terms cannot be copied from the magnetic case.\\

In (\ref{skalres}) the terms in the expansion, which are constant and quadratic in $T$, are just compensated by $K_{02}(T,u_a,4)$, as are terms proportional to $1/\G$ by $f(T,4)/\G$. The finite contribution derives exclusively from $K(T,u_a,4)$ and the second integral and is adjusted to reproduce the correct electron mass renormalization. If one expands the terms in the integrand to higher orders in the fields and performs all the integrations, the contributions from $f(T,4)/\G$ of course cancel the divergencies of $K(T,u_a,4)$ and one can identify in the coefficients the relativistic invariants $F^2$ and $F\tilde{F}$ in polynomials in $a^2$ and $b^2$. The expansion to the fourth power of $e^2$ reads:
\beqn
\bar{\cal L}_{{\mbox{\scriptsize scal}}}^{(2)}[F] &=& \frac{\alpha^3}{\pi m^4} \left( \frac{275(a^4+b^4)+422 a^2 b^2}{2592} \right) \\
& & -\frac{\alpha^4}{m^8} \left( \frac{5159(a^6+b^6)}{16200}+ \frac{8881(a^4b^2+a^2b^4)}{16200} \right) \nonumber \\
& & +\frac{\pi \alpha^5}{m^{12}} \left( \frac{751673(a^8+b^8)}{264600} +\frac{39905(a^2 b^6+a^6 b^2)}{7938} +\frac{323431\ a^4b^4}{56700} \right) +o(e^{10}) \nonumber \\
&=& \frac{\alpha^3}{\pi m^4} \left( \frac{275(\epsilon^2-\eta^2)^2}{2592}+\frac{4(\epsilon \eta)^2}{81} \right) \nonumber \\
& & -\frac{\alpha^4}{m^8} \left( \frac{5159(\epsilon^2-\eta^2)^3}{16200} + \frac{1649(\epsilon^2-\eta^2)(\epsilon \eta)^2}{4050} \right) \nonumber \\
& & +\frac{\pi \alpha^5}{m^{12}} \left( \frac{751673(\epsilon^2-\eta^2)^4}{264600}+\frac{628697(\epsilon^2-\eta^2)^2(\epsilon \eta)^2}{99225}+ \frac{132134(\epsilon \eta)^4}{99225} \right) +o(e^{10}),\nonumber
\eeqn
where we have resubstituted the field strengths from (\ref{epseta}). The terms of first order in $e^2$ coincide exactly with results from conventional QED \cite{RIT,RIT2} and for a purely magnetic field in the limit $\eta \rightarrow 0$ the earlier results \cite{DIMR} of the \wf\ are reproduced. Higher order terms of this expansion are easily extracted from (\ref{skalres}).

\section{Spinor Quantum Electrodynamics}
\label{spin}

We first state the unregularized Lagrangian on the one-loop level from conventional spinor QED \cite{EH,SCH,RIT,RIT2}:
\beqn
{\cal L}_{{\mbox{\scriptsize spin}}}^{\mbox{\scriptsize QED}}[F] &=& {\cal L}_{{\mbox{\scriptsize spin}}}^{(0)}[F] +{\cal L}_{{\mbox{\scriptsize spin}}}^{(1)}[F] \nonumber \\
&\equiv& \frac{\epsilon^2-\eta^2}{2} -\frac{1}{8\pi^2} \int_{0}^{\infty}{\frac{dT}{T^3} e^{-m^2T} \frac{vw}{\tanh(v)\tanh(w)}},
\eeqn
and the dimensional regularized one-loop correction:
\beq
\bar{\cal L}_{{\mbox{\scriptsize spin}}}^{(1)}[F] = -\frac{2}{(4\pi)^{d/2}} \int_{0}^{\infty}{\frac{dT}{T^{d/2+1}} e^{-m_0^2T} \left( \frac{vw}{\tanh(v)\tanh(w)} -\frac{(v^2+w^2)}{3} -1 \right)}.
\eeq
Again we have performed a Wick rotation of the integration variable, otherwise the notation is adopted from chapter \ref{skalar}.\\

In evaluating the expressions of the \wf\ we in general again follow the methods of \cite{DIMR}. The calculation (\ref{determ}) of the functional determinants in the Worldline formulas (\ref{einlo}) and (\ref{zweiloopeinf}) for the two-loop correction to the supersymmetrized Lagrangian results in the following SPT-integral:
\beqn
{\cal L}_{{\mbox{\scriptsize spin}}}^{(2)}[F] &=& \frac{e^2}{(4\pi)^d} \int_{0}^{\infty}{\frac{dT}{T^{d/2+1}}} e^{-m^2T} \int_{0}^{\infty}{d\bar{T}} \int_{0}^{T}{d\tau_a d\tau_b} \int{d\theta_a d\theta_b} \\
& & \times \dt^{-1/2} \left( \frac{\tan(eFT)}{eFT} \right) \dt^{-1/2} \left( \bar{T} -\frac{1}{2} \hat{\cal C}_{ab} \right) \langle -D x_a D x_b \rangle. \nonumber 
\eeqn
The contractions of superderivatives can be expressed in terms of super Green's functions, substituting the Green's function of the scalar case. Instead we immediately write the result in the component field representation, in which the super Green's functions split into modified bosonic and fermionic Green's functions ${\cal G}_{Bab}$ and ${\cal G}_{Fab}$, and use a partial integration in $\tau_a$ to remove second derivative terms and do the Grassmann-integrations as well:
\beqn \label{spinmaster}
{\cal L}_{{\mbox{\scriptsize spin}}}^{(2)}[F] &=& \frac{e^2}{2(4\pi)^d} \int_{0}^{\infty}{\frac{dT}{T^{d/2+1}} e^{-m^2T}} \int_{0}^{\infty}{d\bar{T}} \int_{0}^{T}{d\tau_a} \int_{0}^{T}{d\tau_b} \\
& & \times \dt^{-1/2} \left( \frac{\tan(eFT)}{eFT} \right) \dt^{-1/2} \left( \bar{T} -\frac{1}{2} {\cal C}_{ab} \right) \nonumber \\
& & \times \left( \tr \left( \dot{\cal G}_{Bab} \right) \tr \left( \frac{\dot{\cal G}_{Bab}}{\bar{T}- \frac{1}{2} {\cal C}_{ab}} \right) -\tr \left( {\cal G}_{Fab} \right) \tr \left( \frac{{\cal G}_{Fab}}{\bar{T}- \frac{1}{2} {\cal C}_{ab}} \right) \right. \nonumber \\
& & +\left. \tr \left( \frac{(\dot{\cal G}_{Baa} -\dot{\cal G}_{Bab})(\dot{\cal G}_{Bab} -\dot{\cal G}_{Bbb}+2{\cal G}_{Faa})+{\cal G}_{Fab} {\cal G}_{Fab}-{\cal G}_{Faa} {\cal G}_{Fbb}}{\bar{T}- \frac{1}{2} {\cal C}_{ab}} \right) \right). \nonumber
\eeqn
The modified fermionic Green's function is defined by:
\beqn
{\cal G}_{Fab} \equiv {\cal G}_F (\tau_a,\tau_b) &\equiv & \langle \tau_a \mid \left( \frac{1}{2} \left( \frac{d}{d\tau} -2ieF \right) \right)^{-1} \mid \tau_b \rangle \\
&=& G_F (\tau_a,\tau_b) \frac{e^{-ieFT\g}}{\cos(eFT)}. \nonumber
\eeqn
The usual fermionic Worldline Green's function $G_{Fab}$ is the inverse of $\frac{1}{2} \partial_\tau$ on the circle with antiperiodic boundary conditions:
\beq
G_F (\tau_a,\tau_b) \equiv \sign(\tau_a-\tau_b).
\eeq
Some of the expressions in (\ref{spinmaster}) can already be found in (\ref{dety}) and (\ref{try}), some more are still to be computed. Using (\ref{sincos}) first we find an explicit representation for the fermionic Green's function ${\cal G}_{Fab}$ and its coincidence limit ${\cal G}_{Faa}$:
\beqn
{\cal G}_{Fab} &=& \sign(\tau_a-\tau_b) \left( I_1 \frac{\cosh(v\g)}{\cosh(v)} -i\sigma_1 \frac{\sinh(v\g)}{\cosh(v)}+ \left( 1 \leftrightarrow 2,\ v \leftrightarrow w \right) \right), \nonumber \\
{\cal G}_{Faa} & \equiv & -i \tan(eFT) = -i\sigma_1 \tanh(v) -i\sigma_2 \tanh(w).
\eeqn
Together with (\ref{zwres}) the remaining determinants and traces can be calculated easily and are stated directly in $d=4+\epsilon$ dimensions:
\beqn
\dt^{-1/2} \left( \frac{\tan(eFT)}{eFT} \right) &=& \frac{v}{\tanh(v)} \frac{w}{\tanh(w)}, \\
\pagebreak
\tr \left( {\cal G}_{Fab} \right) \tr \left( \frac{{\cal G}_{Fab}}{\bar{T}-\frac{1}{2} {\cal C}_{ab}} \right) &=& 4 \left( \frac{\cosh(v\g)}{\cosh(v)} + \frac{\cosh(w\g)}{\cosh(w)} + \frac{d-4}{2} \right) \nonumber \\
& & \times \left( \gamma^v \frac{\cosh(v\g)}{\cosh(v)} +\gamma^w \frac{\cosh(w\g)}{\cosh(w)} + \gamma \frac{d-4}{2} \right), \nonumber \\
\tr \left( \frac{2{\cal G}_{Faa}(\dot{\cal G}_{Baa}-\dot{\cal G}_{Bab})}{\bar{T}-\frac{1}{2}{\cal C}_{ab}} \right) &=& 4\left( \gamma^v \left( 1-\frac{\cosh(v\g)}{\cosh(v)} \right) + \gamma^w \left( 1-\frac{\cosh(w\g)}{\cosh(w)} \right) \right), \nonumber \\
\tr \left( \frac{{\cal G}_{Fab} {\cal G}_{Fab}-{\cal G}_{Faa} {\cal G}_{Fbb}}{\bar{T}-\frac{1}{2}{\cal C}_{ab}} \right) &=& 2 \left( \gamma^v \frac{\cosh^2(v\g)+\sinh^2(v\g)-\sinh^2(v)}{\cosh^2(v)} \right. \nonumber \\ 
& & + \left. \gamma^w \frac{\cosh^2(w\g)+\sinh^2(w\g)-\sinh^2(w)}{\cosh^2(w)} +\gamma \frac{d-4}{2} \right). \nonumber
\eeqn
Inserting into (\ref{spinmaster}), we can arrange the integrand in a form, in which it explicitly only depends on the SPT-variables, respectively the Green's functions. As in the scalar case we rescale the $\bar{T}$-integration to the unit cirle and employ translation invariance of the zero-point to set $\tau_b=0$, which implies $\sign(\tau_a-\tau_b)=1$, so that (\ref{greenfu}) is also satisfied again. We then introduce an integrand $J(T,\bar{T},u_a,d)$ in accordance with
\beq
{\cal L}_{{\mbox{\scriptsize spin}}}^{(2)}[F]= \frac{e^2}{(4\pi)^d} \int_{0}^{\infty}{\frac{dT}{T^{d-1}} e^{-m^2T}} \int_{0}^{\infty}{d\bar{T}} \int_{0}^{1}{du_a J(T,\bar{T},u_a,d)},
\eeq
which we expand under the integral in powers of $T$ and $\G$. To identify the subtraction terms, integration and differentiation are interchanged and the $\bar{T}$-integration is performed coefficient by coefficient:
\beq
L(T,u_a,d) \equiv \int_{0}^{\infty}{d\bar{T}\ J(T,\bar{T},u_a,d)} =L_{02}(T,u_a,d)+g(T,d)\G^{1-d/2}+ o\left( T^4,\G^{2-d/2}\right).
\eeq
In contrast to scalar QED right from the beginning no terms are appearing that were proportional to $\G^{-d/2}$, and the Maxwell term is of the order $\epsilon^0$, so that the Lagrangian from (\ref{spinmaster}) can be regularized analogously to (\ref{linkomb}) without any further modification. For the divergent coefficients we find:
\beqn
L_{02}(T,u_a,d) &=& -4(d-1)\G^{1-d/2} -\frac{v^2+w^2}{3d} \\
& & \times \left( 4(d-1)(d-4)\G^{1-d/2} +4(d-2)(d-7)\G^{2-d/2} \right), \nonumber \\
g(T,d) &=& -\frac{4(d-1)}{3d} \left( \left( \frac{vw}{\tanh(v) \tanh(w)} \right) \left( 6v (\coth(v)-\tanh(v)) \right. \right. \nonumber \\
& & +\left. \left. 6w(\coth(w)-\tanh(w)) +3(d-4) \right) -(d-4)(v^2+w^2)-3d \right). \nonumber
\eeqn
Because of $g(T,d)=o(T^4)$ the divergencies are completely seperated. To obey the on shell subtraction scheme, i.e. to choose the finite subtraction term appropriate to have (\ref{onshbed}) satisfied by $\delta {\cal L}_{{\mbox{\scriptsize spin}}}^{(2)}[F]$ as well, we relate $g(T,d)$ to $\bar{\cal L}_{{\mbox{\scriptsize spin}}}^{(1)}[F]$ again by a partial integration:
\beq
g(T,d)= \frac{8(d-1)}{d} T^{d/2+1} \frac{d}{dT} \left( T^{-d/2} \left( \frac{vw}{\tanh(v)\tanh(w)} -\frac{v^2+w^2}{3}-1 \right) \right).
\eeq
Using further
\beq
\delta m^2_0= \frac{\alpha m_0^2}{4\pi} \left( -\frac{6}{\epsilon}+4-3(\gamma-\ln(4\pi))-3\ln(m_0^2) \right) +o(\epsilon)
\eeq
with (\ref{geps}), one now confirms by partial integration of the derivative term in $g(T,d)$ that:
\beqn
\delta m_0^2 \frac{\partial \bar{\cal L}_{{\mbox{\scriptsize spin}}}^{(1)}[F]}{\partial m_0^2} &=&
\frac{\alpha}{(4\pi)^3} \int_{0}^{\infty}{\frac{dT}{T^{d-1}} e^{-m_0^2T}} \int_{0}^{1}{du_a \ g(T,d) \ \G^{1-d/2}} \\
& & +\frac{\alpha m_0^2}{(4\pi)^3} \int_{0}^{\infty}{\frac{dT}{T^2} e^{-m_0^2T}} \left( \frac{vw}{\tanh(v) \tanh(w)} -\frac{v^2+w^2}{3}-1 \right) \nonumber \\ 
& & \times \left( 18-12\gamma -12\ln(m_0^2T)+\frac{12}{m_0^2T} \right). \nonumber 
\eeqn
Finally we obtain the result for the on shell renormalized two-loop correction to the Euler-Heisenberg Lagrangian in spinor QED:
\beqn \label{spinres}
\bar{\cal L}_{{\mbox{\scriptsize spin}}}^{(2)}[F] &=& \frac{\alpha}{(4\pi)^3} \int_{0}^{\infty}{\frac{dT}{T^3} e^{-m_0^2T}} \int_{0}^{1}{du_a} \left( L(T,u_a,4)-L_{02}(T,u_a,4) \right) -\delta m_0^2\frac{\partial \bar{\cal L}_{{\mbox{\scriptsize spin}}}^{(1)}[F]}{\partial m_0^2} \nonumber \\
&=& \frac{\alpha}{(4\pi)^3} \int_{0}^{\infty}{\frac{dT}{T^3} e^{-m^2T}} \int_{0}^{1}{du_a} \left( L(T,u_a,4)-L_{02}(T,u_a,4)-g(T,4) \G^{-1} \right) \nonumber \\
& & -\frac{\alpha}{(4\pi)^3} \int_{0}^{\infty}{\frac{dT}{T^2} e^{-m^2T}} \left( \frac{e^2abT^2}{\tanh(eaT) \tanh(ebT)} -\frac{e^2(a^2+b^2)T^2}{3} -1 \right) \nonumber \\
& & \times \left( 18-12\gamma -12\ln(m^2T)+\frac{12}{m^2T} \right).
\eeqn
The $\bar{T}$-integration of $J(T,\bar{T},u_a,d)$ can again be performed elementarily for finite values of the other variables and the remaining parameter integral stays finite. The cancellations of the divergent parts of the integral over $L(T,u_a,4)$ occurr exactly in the same manner as in the scalar expression and the second integral adjusts the electron mass renormalization. We also find structural similarities to the results for a purely magnetic background field. The expansion of the integrand to the fourth order in $e^2$ leads to:
\beqn
\bar{\cal L}_{{\mbox{\scriptsize spin}}}^{(2)}[F]&=& \frac{\alpha^3}{\pi m^4} \left( \frac{16(\epsilon^2-\eta^2)^2}{81}+\frac{263(\epsilon \eta)^2}{162} \right) \\
& & - \frac{\alpha^4}{m^8} \left( \frac{1219(\epsilon^2-\eta^2)^3}{2025} +\frac{8656(\epsilon^2-\eta^2)(\epsilon \eta)^2}{2025} \right) \nonumber \\
& & +\frac{\pi \alpha^5}{m^{12}} \left( \frac{541232(\epsilon^2-\eta^2)^4}{99225} + \frac{470912(\epsilon^2-\eta^2)^2(\epsilon \eta)^2}{11025} +\frac{3815584(\epsilon \eta)^4}{99225} \right) +o\left( e^{10} \right). \nonumber
\eeqn
It is very easy to obtain higher orders from the given formulas. The lowest order in $e^2$ is perfectly matching the results from standard spinor QED\cite{RIT,RIT2}. Of course, the expansion reproduces earlier Worldline results for the purely magnetic field \cite{DIMR} in the limit $\eta \rightarrow 0$.\\

\vspace{1cm}

\hspace{-.65cm} {\bf{Acknowledgements}}\\
We would like to thank C. Schubert for helpful remarks.

\pagebreak

\end{document}